\documentclass[pre,twocolumn,a4paper,showpacs]{revtex4-1}
\usepackage{amssymb}
\usepackage{graphicx}
\begin{document}
\title{Social interaction as a heuristic for combinatorial optimization problems}

\author{Jos\'e F. Fontanari}
\affiliation{Instituto de F\'{\i}sica de S\~ao Carlos,
  Universidade de S\~ao Paulo,
  Caixa Postal 369, 13560-970 S\~ao Carlos, S\~ao Paulo, Brazil}

\begin{abstract}

We investigate the performance of a variant of Axelrod's model for dissemination of culture -- the  Adaptive Culture Heuristic (ACH) -- on
solving  an NP-Complete optimization problem, namely, the  classification of  binary input patterns of size $F$ by a Boolean Binary Perceptron. In this heuristic,
$N$ agents, characterized by binary strings of length $F$  which represent possible solutions to
the optimization problem, are fixed at the sites of a square lattice and interact  with their nearest neighbors only. The interactions are such that the  agents' strings (or cultures) become more similar to the low-cost strings of  their neighbors resulting in the dissemination  of these  strings across the lattice. 
Eventually  the dynamics freezes into a homogeneous absorbing configuration in which all agents exhibit identical solutions to the optimization problem.
We find through extensive simulations  that the probability 
of finding the optimal solution is a function of the reduced variable $F/N^{1/4}$ so that the number of agents must increase with the
fourth power of the problem size, $N \propto F^ 4$, to guarantee   a fixed probability of success. In this case, we find that the  relaxation time to reach an absorbing configuration scales with $F^ 6$ which can be interpreted
as the overall computational cost of the ACH to find an optimal set of  weights  for a Boolean Binary Perceptron,
given a fixed  probability of success.

\end{abstract}

\pacs{87.23.Ge 89.75.Da, 89.70.Eg, 05.50.+q}
% Uncomment for Submitted to journal title message
% Comment out if separate title page not required
\maketitle
%%%%%%%%%%%%%%%%%%%%%%%%%%%%%%%%%%%%%%%%%%%%%%%%%%%%%%%%%%%%

%
\section{Introduction} \label{sec:Intro}

In the early eighties, the perception that the dynamics of the celebrated  Hopfield model of associative memory \cite{Hopfield_82} 
was solving an  optimization problem, namely, that  of finding  which stored pattern is closest to the input configuration,  led to the proposal of a powerful 
general-purpose optimization heuristic,  the so-called Hopfield-Tank
neural network \cite{Hopfield_85}. A similar situation happened in the late nineties, when Kennedy \cite{Kennedy_98} pointed out that 
Axelrod's model of culture dissemination \cite{Axelrod_97} could work as a collective problem-solving system provided that one associates the
cultures of the agents (represented by  strings of integer numbers) with the trial solutions of a given optimization problem. That proof-of-concept paper demonstrated  then that social interaction is a natural computation method.

In contrast with Hopfield-Tank neural network, the optimization heuristic based on social interaction, which henceforth we  refer to as 
the  Adaptive Culture Heuristic (ACH), has not enjoyed great popularity among the physics and computer science community, 
perhaps because of the appearance at the same time  of a related algorithm, called particle swarm optimization, which has by now 
become an established optimization paradigm \cite{Swarm,Eberhart_01}. Particle swarm  optimization, however, suits best to search in 
space of real-valued variables, whereas ACH is proper to explore configuration spaces of discrete-valued variables, which is the case of  most combinatorial optimization problems that have attracted the attention of the 
statistical physics community \cite{MPV_86}. Here we attempt to change this situation by showing that the performance of the ACH  seems to scale very favorably  (it improves exponentially fast)  with the number of agents in the system.

Following Axelrod's model \cite{Axelrod_97}, the ACH  requires a population of $N = L^2$  agents placed at the sites of a square lattice of size $L \times L$ with periodic boundary conditions. The agents  can interact  with their four nearest neighbors only. Each agent is characterized by a binary string  of length $F$, which represents  the agent's solution to the optimization problem  in the ACH interpretation.  In  Axelrod's model this string, which is not necessarily binary,  represents the culture of the agent. The interaction  between any two neighboring agents occurs whenever the agents have different strings, regardless of their associated cost,  and it is such that  the string of the agent with the higher cost solution is slightly modified to become more similar to that of the more efficient partner. 

We recall that in Axelrod's model the  interaction  between two neighboring agents  takes place with probability proportional  to the number of  entries their cultural strings have in common and so agents with completely different cultures do not interact. In the case the agents are allowed to interact, the interaction results in  the increase of the similarity between  the cultures of the two agents, as in the ACH update rule. The fact that some agents are prohibited to interact is the key ingredient for the existence of stable  globally polarized 
states (i.e., culturally heterogeneous  absorbing configurations) which is the major outcome of Axelrod's model  \cite{Axelrod_97}. In the ACH, however, we seek homogeneous absorbing configurations associated to low cost solutions of the target optimization problem and so   the homogenizing interactions are always allowed regardless of the similarity between the strings of the neighboring agents \cite{Kennedy_98}.

In order to obtain statistically  reliable results on the scaling of the performance of  ACH with the size $F$ of the optimization problem  and the number $N$ of agents in the lattice, we focus on a specific optimization problem which involves  the manipulation of binary variables only, namely, the categorization of binary  input patterns by the Boolean Binary Perceptron. This is a NP-Complete problem \cite{Garey_79} for which there is no efficient specific heuristic  optimization method available \cite{FM_91} and whose random version has  received a considerable attention from the statistical mechanics community (see, e.g., \cite{Gardner_89,KM_89,FK_90,Gyorgyi_90,Seung_92,FM_93}) because its phase diagram  exhibits a frozen phase  similar to that of the Random Energy Model \cite{Derrida_81}.

The main result of this paper is that,  given a fixed probability of success,
the overall computational cost of ACH to find a minimum-cost solution for the learning problem in a Boolean Binary Perceptron   scales with the sixth power of the size of the input string. Of course, this finding has no implication on the celebrated $NP \neq P$ conjecture of computer science since the $F^6$  scaling holds for typical realizations of the input-output mapping, rather than for all  realizations as would be required to disprove that assertion. In addition,  ACH is {\it not} a deterministic algorithm which disqualifies the heuristic as a candidate to disprove the $NP \neq P$  conjecture.

The rest of this paper is organized as follows. First   we introduce the target optimization problem  -- categorization of binary patterns by the Boolean Binary Perceptron --  on which we will measure the performance of the Adaptive Culture Heuristic (Sect.~\ref{sec:BBP}). This heuristic is then described in great detail   in Sect.~ \ref{sec:ACH} and the results of its performance on
the training task, measured by the probability that the heuristic finds a minimum cost solution,  are presented in Sect.~ \ref{sec:res}. In this section we present also the
performance of the ACH in the case the  agents are placed at the nodes of random symmetric graphs and argue that the  square lattice connectivity $C=4$ yields the best performance.
Finally, in Sect.~ \ref{sec:conc} we present our concluding remarks.

\section{The Boolean Binary Perceptron}\label{sec:BBP}

The Boolean Binary Perceptron is a single-layer neural network whose weights are constrained to take on binary values only.
More explicitly, the network consists of an input layer with $F$ binary neurons $s_k = \pm 1, k=1,\ldots,F$ with each input neuron  connected
to the output unit $o = \pm 1$ through the weights $w_k = \pm 1,  k=1,\ldots,F$. The state of the output unit is given by the equation
\begin{equation}\label{o}
 o = \mbox{sign} \left ( \sum_{k=1}^F w_k s_k \right )
\end{equation}
where $\mbox{sign} \left ( x \right ) = 1 $ for $ x \geq 0$ and $-1$ otherwise. We will restrict $F$ to take on 
odd integer values only, so we can guarantee that the argument of the sign function will never vanish. The learning task is
to find a set of  weights $\hat{w}_k = \pm 1,  k=1,\ldots,F$ that emulates the input-output mapping 
$ \left ( s_1^l, \ldots, s_F^l \right ) \to t^l $
for $l = 1, \ldots, M$. If the weights were allowed to assume real values then this learning task could easily be accomplished by the
perceptron learning algorithm or by the Widrow-Hoff rule \cite{Duda_73}. However, when the binary constraint is taken into account
the learning task becomes an NP-complete problem since it is equivalent to integer programming \cite{Garey_79}. Assuming that
$NP \neq P$, this means that no deterministic algorithm can find  $\hat{w}_k,  k=1,\ldots,F$ (if  it
exists) for any realization of the input-output mapping in a time that grows polynomially
with the parameter $F$.

Here we focus on random versions of the input-output mapping where the input entries $s_k^l$ are 
statistically
independent random variables chosen as $\pm 1$ with equal probability. As for the output $t^l$ we consider  two schemes.
In the first scheme, we choose $t^l = \pm 1$ at random with equal probability -- so-called random output mapping. In this case, it is not possible to guarantee
that there is a set of binary weights that emulates the input-output mapping  perfectly. In fact, statistical mechanics studies based on the 
landmarking paper by Gardner \cite{Gardner_88}, show that in the limit $F \to \infty$ there are optimal sets of weights provided that
the ratio $\alpha \equiv M/F$ is less than $\alpha_c^r \approx 0.83$ \cite{KM_89,FM_93}. So, in this limit, we say that the input-output
mapping is linearly Boolean separable for $\alpha < \alpha_c^r$. 

However, it is convenient to consider input-output mappings which are linearly 
Boolean separable for any choice of the parameters $F$ and $M$. This observation motivates the second scheme to set the values of the outputs $t^l$ , which are given by
\begin{equation}\label{LBSP}
t^l = \mbox{sign} \left ( \sum_{k=1}^F w_k^0 s_k^l \right )
\end{equation}
for $l=1, \ldots, M$. Here $w_k^0, k=1, \ldots, F$  are statistically independent random variables that take on the 
values $\pm 1$ with equal probability. Clearly, such input-output mapping is linearly Boolean separable by construction, since the
set of binary weights $w_k^0, k=1, \ldots, F$ emulates it perfectly. The solution weight space of this problem was studied numerically
\cite{Gardner_89,FK_90} and analytically \cite{Gyorgyi_90,Seung_92}, resulting in the conclusion that   in the limit $F \to \infty$
the only solution to the mapping is the teacher perceptron  $w_k^0, k=1, \ldots, F$ for $\alpha > \alpha_c^{0} \approx 1.245$.

From the perspective of interpreting  the neural network training as an optimization problem we define the following
cost function
\begin{equation}\label{cost}
E \left ( \{ w_k \} \right ) = \sum_{l=1}^M \Theta \left ( - t^l \sum_{k=1}^F w_k s_k^l \right )
\end{equation}
where $\Theta \left ( x \right ) = 1$ if $x \geq 0$ and $0$ otherwise. Hence the cost $E$ yields the number of
misclassified inputs and so  its minimum (optimum) value  is zero in
the case of a linearly Boolean separable mapping. 

In this paper we will concentrate mostly on the 
linearly Boolean separable mappings defined by Eq.\ (\ref{LBSP}) because in this case the optimal solution is known {\it a priori}
so we can evaluate the performance of the ACH for relatively large problems ($F < 200$), whereas in the random
output mapping  we are restricted to the range $F < 25$, since we need to carry out an exhaustive search over  the $2^F$ possible 
weight configurations in order to find the  minimum cost solution. However, our findings indicate that, regarding the scaling with  respect to the relevant parameters of the problem,  the  performance of 
the heuristic is essentially the same regardless of whether the mapping is linearly Boolean separable or not.

\section{The Adaptive Culture Heuristic}\label{sec:ACH}

 The set of weights of a Boolean Binary Perceptron is completely specified by a binary string of length $F$. In the adaptive culture heuristic,
each such string  is interpreted as the culture of an agent and its cost, given by Eq.\ (\ref{cost}), measures the unworthiness
of the culture. The idea behind the ACH  is that the agents should prefer to adopt more valuable cultures, i.e., those cultures associated with low cost values \cite{Kennedy_98}. In this context, it is more convenient to refer to  the strings that characterize the agents  as solutions rather than cultures.

As already pointed out, the agents
are fixed at the sites of a square lattice of size 
$L \times L$ with periodic boundary conditions and can interact with their four nearest neighbors only. 
At each time we pick an agent at random 
(this is the target agent) as well as one of its four neighbors.  These two agents will interact provided that the cost (\ref{cost})  of the solution associated to 
the target agent is greater or equal to the cost of the solution associated to  the randomly selected  neighbor. 
An interaction consists of selecting at random and then flipping one of the entries which distinguish the target agent 
from its neighbor. Note that only the string  of the target agent is updated, i.e., the agent with the higher cost solution  is changed to become more similar to its neighbor. This change may actually increase the cost of the solution of the target agent, due to the highly nonlinear dependence of the cost (\ref{cost})  on the individual entries of the  binary string.
This procedure is repeated until  the dynamics  freezes in a  homogeneous  absorbing configuration.  We can guarantee that the frozen configurations are homogeneous because we allow interactions, and so changes in the target agent,  even  when the two interacting agents have the same cost value. 

Because of the need to re-calculate the cost function after each interaction, the implementation of the ACH to search for near optimal weights  of the Boolean Binary Perceptron is a very
computationally  demanding problem and so an extensive statistical analysis of the performance of this heuristic requires a highly optimized code. In particular,
to simulate efficiently the ACH for large lattices we use a procedure based on the concept of active agents (see \cite{Barbosa_09,Peres_10}).  An active agent is an agent whose solution  differs from the solution  of at least one of its four neighbors. 
Clearly, only active agents can change their strings and so it is more efficient  to select the
target agent randomly from the list of active agents rather than from the entire lattice.  In the case that
the solution string of the target agent is modified by the updating rule, we 
need to re-examine the active/inactive status of the target agent as well as of all its neighbors so as
to update the list of active agents.  The dynamics is frozen when the list of active agents is empty.  Note that the cost of the solution string plays no role in the definition of active agents. 

\section{Results}\label{sec:res}

All our results are obtained for $M=2F$ so that for the linearly Boolean separable case  the teacher set of weights $w_k^ 0$ is
the only global minimum (zero-cost) solution of the cost function (\ref{cost}), provided that $F$ is sufficiently large. However, what is crucial  for
our purposes is the knowledge  that for any value of $F$ there is at least one solution for which the cost is zero, so that we can focus on the
number of runs of the ACH which results in this minimal cost, regardless of whether the actual solution found by the heuristic is the teacher solution or another
degenerate  zero-cost solution. In particular, for each realization of the input-output mapping we run the ACH for $100$  random initial settings
of the agents'  solutions and calculate the fraction of runs for which  the heuristic reaches a minimum cost solution. This fraction is then averaged over
a variable number, ranging from $500$ to $10^6$, of realizations of the input-output mapping. 

\begin{figure}
\includegraphics[width=0.52\textwidth]{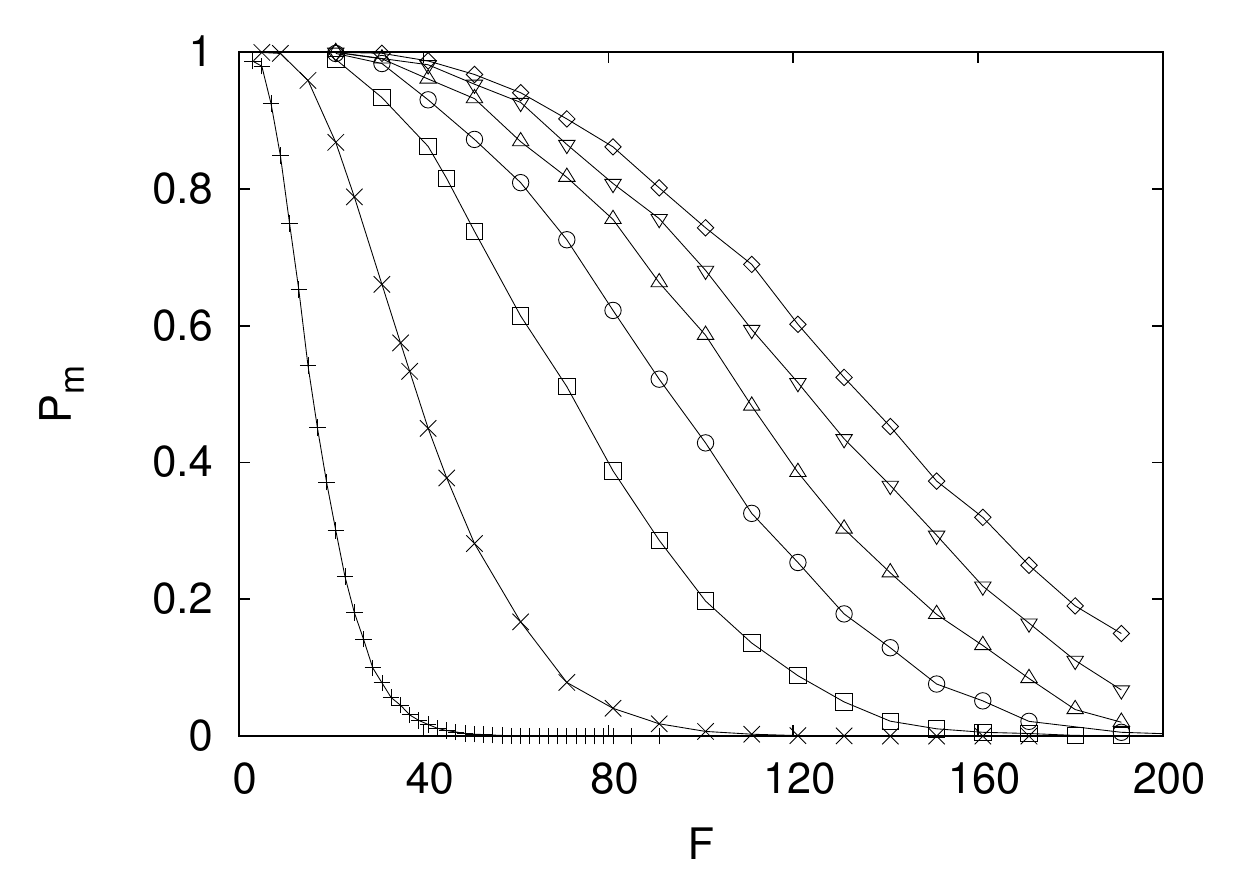}
%\centerline{\epsfig{width=0.52\textwidth,file=fig1.eps}}
\par  
\caption{The probability that a run of the ACH finds a zero-cost solution for linearly Boolean separable mappings as function of
the input size $F$ for lattices with (left to right)  $N = 5^ 2 , 10^ 2, 20^ 2, 30^ 2, 40^ 2, 50^ 2$ and $60^ 2$ agents.  The error bars are smaller than the sizes of the symbols and the lines are guides to eye.
\label{fig:1} }
\end{figure}

As pointed out before, most of our results are for the  linearly Boolean separable case since in this case we know by construction the cost
of the optimum solution and so we can study  the performance of the heuristic for large values of $F$.  At the end of this section we present some results for the random Boolean mapping in the region $F \leq 25$ since then we  first need to perform an exhaustive search in the solution space to find the minimum cost. The main quantity we focus here is
the  mean fraction of runs for which the heuristic reached the minimum-cost solution, which  can be interpreted as the  probability $P_m$  that a run of the  ACH 
finds the optimum cost. This quantity is shown in Fig.\ \ref{fig:1} for the linearly Boolean separable case as function of the size $F$ of the problem  and  of the number $N$ of agents in the system.

%----------------------------------------------------------------------------------------------------
%----------------------------------------------------------------------------------------------------

\begin{figure}
\includegraphics[width=0.52\textwidth]{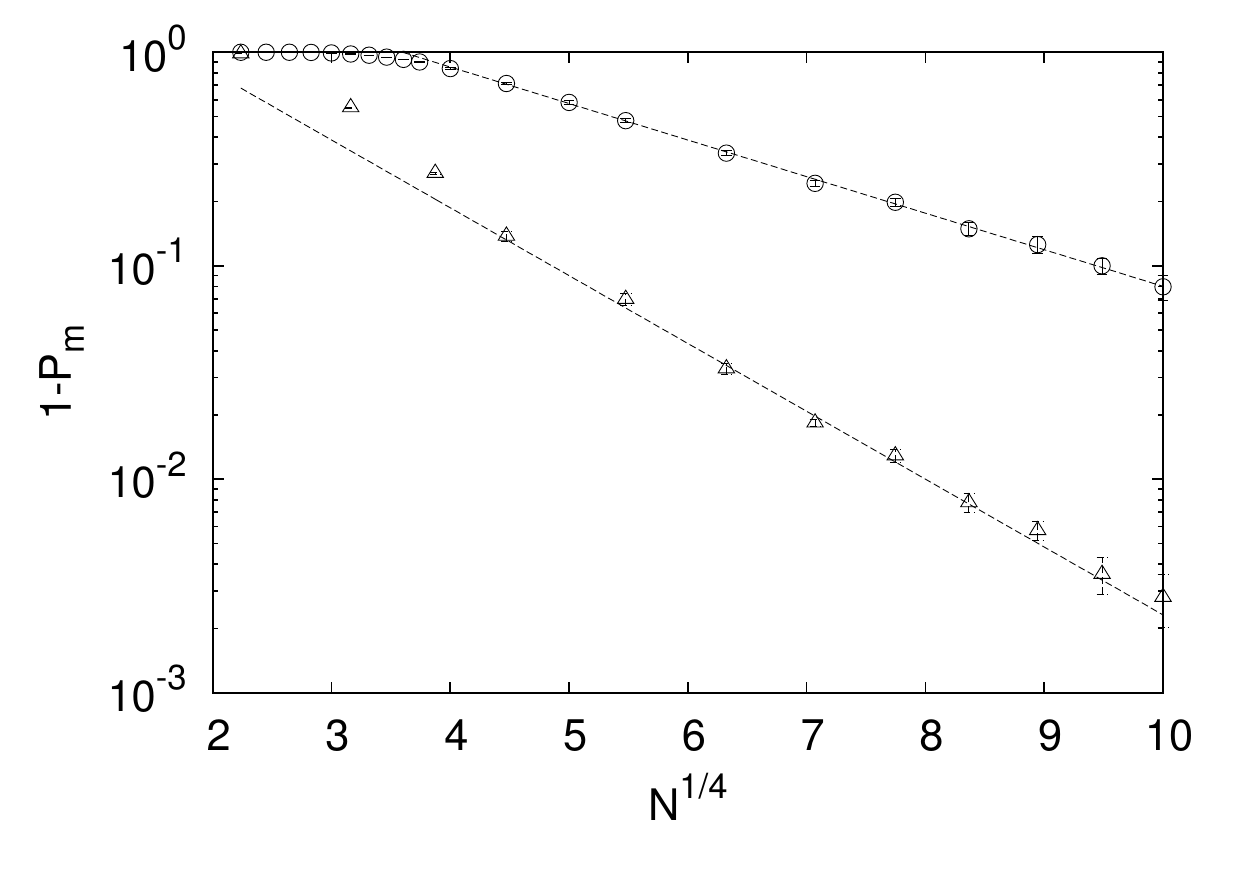}
%\centerline{\epsfig{width=0.52\textwidth,file=fig2.eps}}
\par  
\caption{Semi-logarithmic  plot of the probability  $1-P_m$ that a run of the ACH does not find a zero-cost solution for linearly Boolean separable mappings  as function of $N^{1/4}$ for  $F=91 \left ( \bigcirc \right )$ and
$F=41 \left ( \triangle \right )$.    The dashed straight lines are the fittings $1 -P_m = b_F \exp \left (- a_F N^{1/4} \right )$.
\label{fig:2} }
\end{figure}

Figure \ref{fig:1} reveals a most surprising aspect about the performance of the ACH, namely, that for small $N$, say $N=5^2$, a  fourfold increment on the number of agents in the system, increases the probability of finding an optimal solution by several orders of magnitude. Actually,  this observation holds true even for large $N$,  provided that $F$ is large enough. To quantify this observation, in Fig.\ \ref{fig:2} we show how $P_m$ approaches $1$ as the number of agents  $N$ increases for  two values of the input size $F$. This analysis  shows that for $N > 30^2$,  the probability $1- P_m$  that the heuristic fails to find the optimum cost vanishes like $\exp \left ( -a_F N^{1/4} \right )$  where the (fitting) parameter $a_F$   is inversely proportional to  $F$.

%----------------------------------------------------------------------------------------------------
%----------------------------------------------------------------------------------------------------

%----------------------------------------------------------------------------------------------------
\begin{figure}
\includegraphics[width=0.52\textwidth]{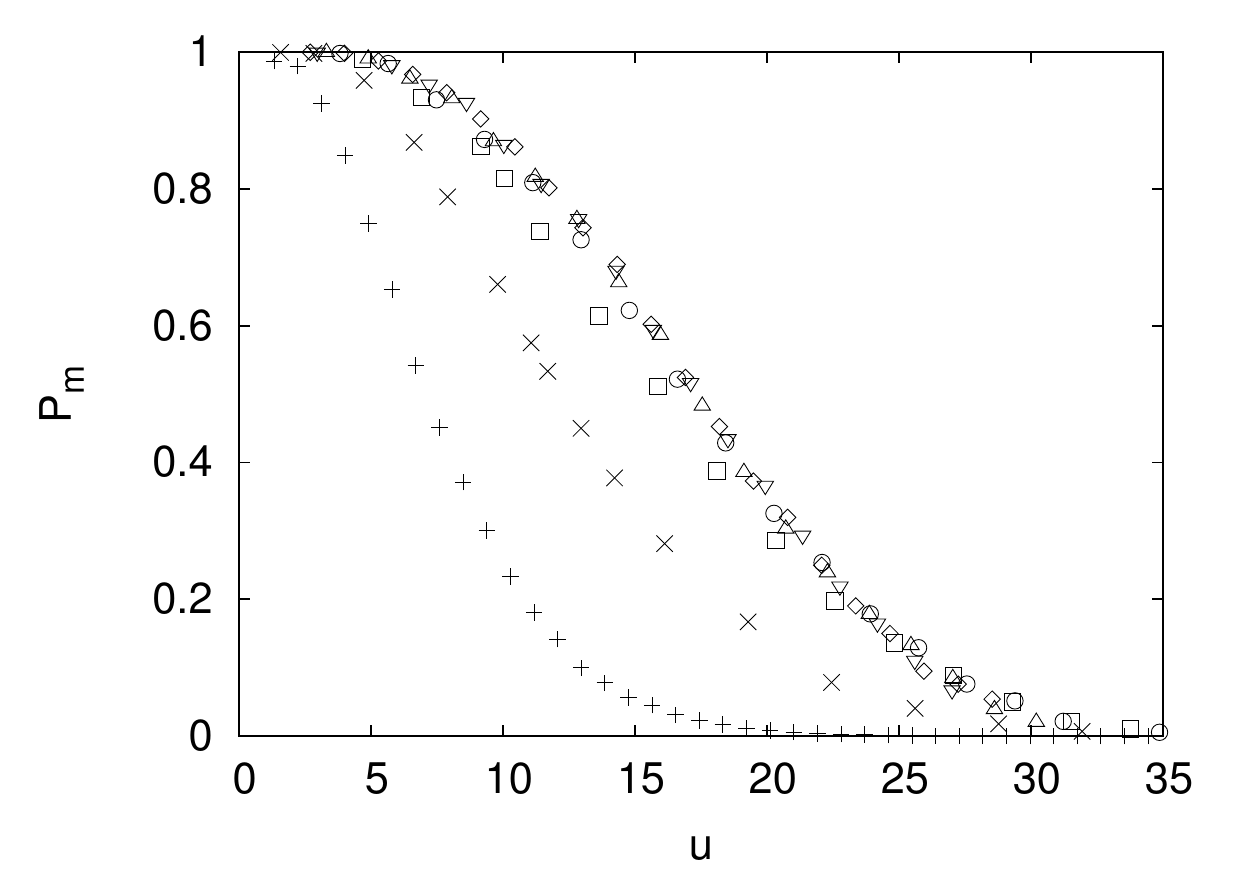}
%\centerline{\epsfig{width=0.52\textwidth,file=fig3.eps}}
\par  
\caption{The same data exhibited in Fig.\ \ref{fig:1} plotted in terms of the rescaled variable 
$u \equiv = F/N^{1/4}$. The data for $N \geq 30^2$ lie in approximately the same curve 
given by the scaling function $P_m = g \left (u \right )$.
\label{fig:3} }
\end{figure}
%----------------------------------------------------------------------------------------------------

%----------------------------------------------------------------------------------------------------
\begin{figure}
\includegraphics[width=0.52\textwidth]{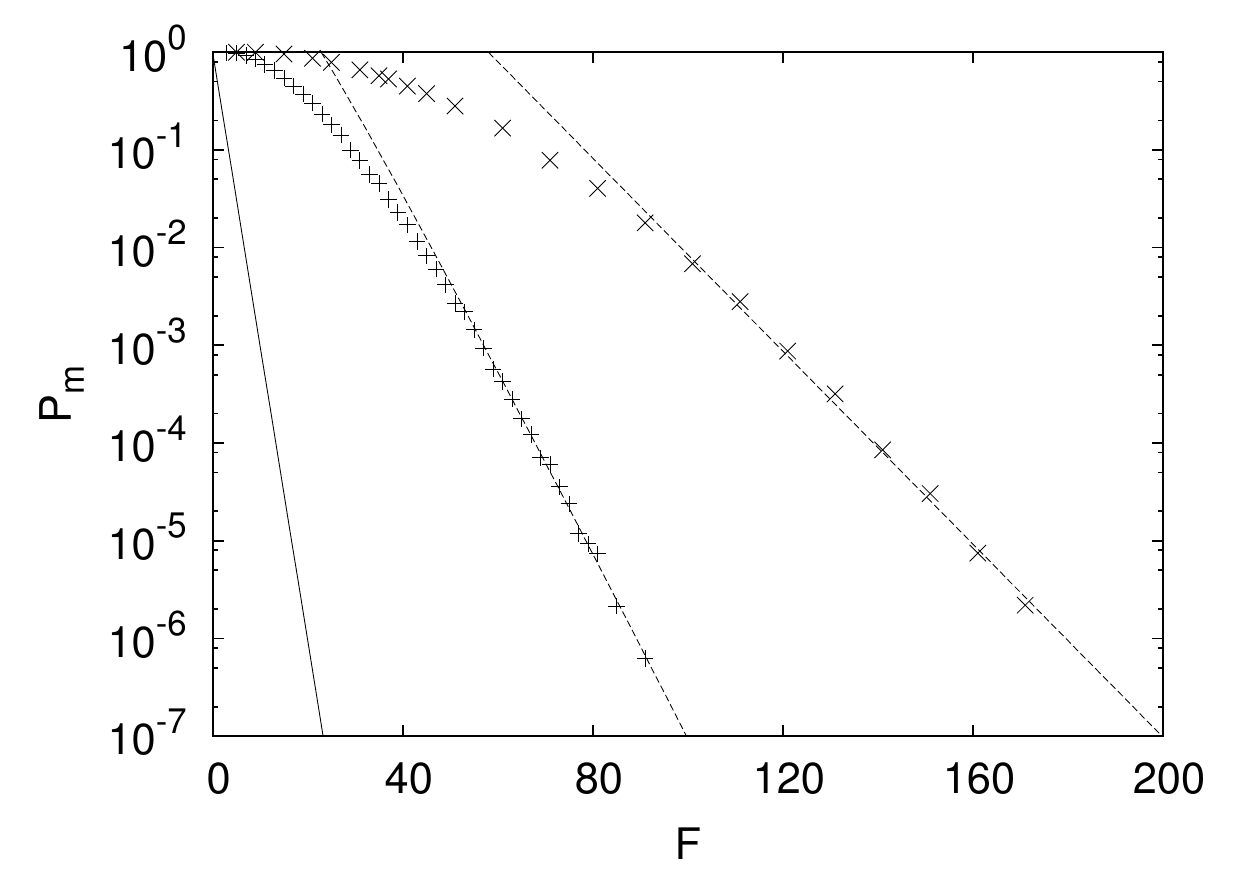}
%\centerline{\epsfig{width=0.52\textwidth,file=fig4.eps}}
\par  
\caption{Semi-logarithmic  plot of the probability that a run of the ACH finds a zero-cost solution for linearly Boolean separable mappings as function of the input size $F$ for 
$N=5^2 (+)$ and  $N=10^2 (\times)$.  The error bars are smaller than the sizes of the symbols. The solid straight line yields the probability that the optimal solution is chosen in a random  selection,  $2^{-F}$, whereas 
 the dashed straight lines are the fittings $P_m = b_N \exp \left (- a_N F \right )$.
\label{fig:4} }
\end{figure}
%----------------------------------------------------------------------------------------------------

These findings  prompt us to redraw Fig.\ \ref{fig:1} in terms of  the rescaled variable $u \equiv F/N^{1/4}$, which is  done in Fig.\ \ref{fig:3}. The collapse of the data for $N \geq 30^2$ into a single curve implies that 
$P_m = g \left ( u \right )$.  We note that the failure of the scaling function $g  \left ( u \right )$  to describe the data for $N < 30^2$  was already expected from the results of Fig.\ \ref{fig:2}. In fact, those results show that in the limit $u \to 0$ we have $g \left ( u \right ) \sim \exp \left ( - a/u \right )$ with $a \approx 0.5$. 

The study of the scaling function  $g \left ( u \right )$ in the  other extreme limit,  $u \to \infty$, requires very large input sizes  ($F > 200$) for 
relatively large lattices ($N \geq 30^2$) which is computationally unfeasible because of the need to use
a huge number of samples to get a reliable statistics  since  $P_m \to 0$  in this limit. Nevertheless, in Fig.\ \ref{fig:4} we present such analysis in the case of small lattices  $N=5^2$ and $N=10^2$, for which we know
the scaling behavior is not valid.  As expected, the results show that $P_m$ vanishes exponentially with increasing $F$, i.e., $P_m \sim \exp \left (- a_N F \right )$. Here the fitting parameter is given by  $a_N \approx 1/N^{1/2}$, indicating that for small $N$ the gain on performance obtained by increasing the number of agents is much larger than the gain in the scaling regime where $a_N \sim 1/N^{1/4}$. In addition, Fig.\ \ref{fig:4} is useful to highlight the enormous gain on performance resulting from the increase of the number of agents involved in the optimization procedure.

\begin{figure}
\includegraphics[width=0.52\textwidth]{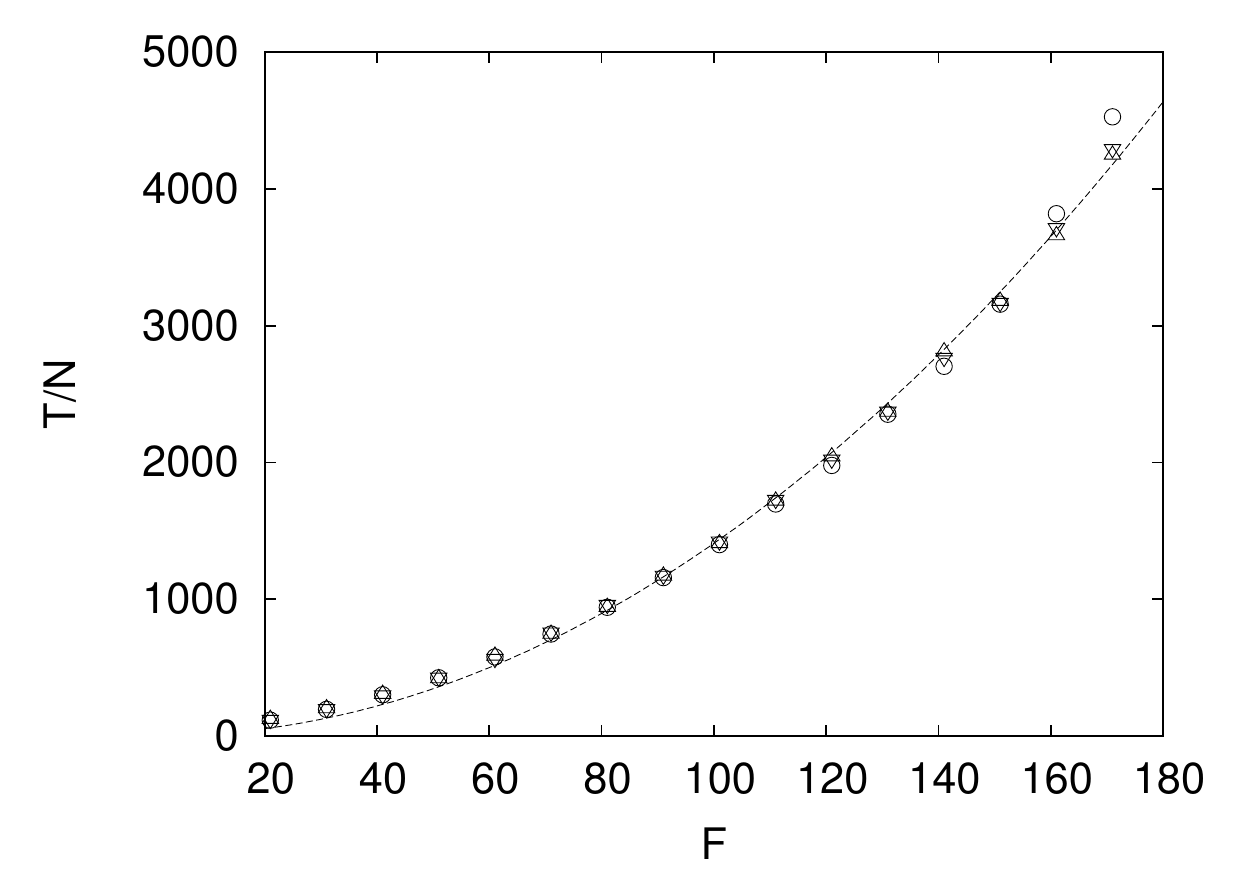}
%\centerline{\epsfig{width=0.52\textwidth,file=fig5.eps}}
\par  
\caption{Scaled relaxation time of the ACH as function of the input size $F$ for 
$N=30^{2} (\bigcirc), 40^2 (\triangle) $ and $50^2 (\bigtriangledown)$.  The error bars are smaller than the sizes of the symbols. The dashed curve is the  
fitting $T/N =  0.12 F^2 $.
\label{fig:5} }
\end{figure}

A most appealing feature of the ACH is that the dynamics always freezes in a homogeneous absorbing configuration and so the algorithm halts.  We must note, however, that the ACH is a stochastic heuristic since the same initial configuration of the lattice can lead to different  absorbing configurations depending on the sequence of site updates. The fact that the  dynamics  eventually freezes allows us to define a relaxation time for the ACH, which is a quite  unexpected bonus for a stochastic heuristic. Accordingly, in Fig.\ \ref{fig:5} we show the scaled average
relaxation  time $T/N$  as function of the input size $F$. The  unsurprising  fact that $T$ scales linearly with the number of agents $N$ is manifested by the coincidence of results for different lattice sizes. The instructive result here is that $T$ grows   with the square of the input size only. This result will be useful for the  evaluation of the overall computational  demand of the ACH (see Sect. \ref{sec:conc}).

%----------------------------------------------------------------------------------------------------
%----------------------------------------------------------------------------------------------------

\begin{figure}
\includegraphics[width=0.52\textwidth]{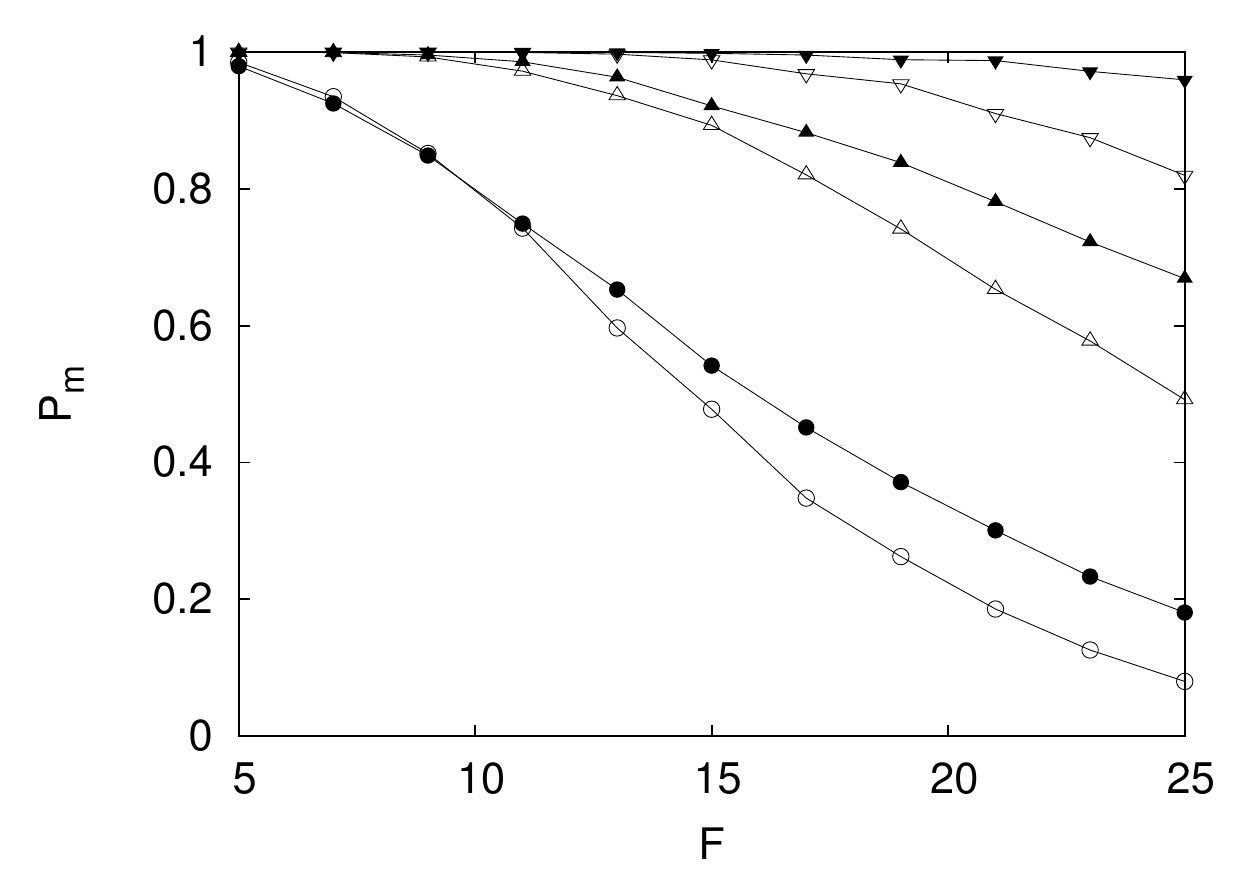}
%\centerline{\epsfig{width=0.52\textwidth,file=fig6.eps}}
\par  
\caption{Comparison between the performances of the ACH for the random mapping (open symbols) and 
the linearly Boolean separable mapping (filled symbols) for $N= 5^2 (\bigcirc), 10^2 (\triangle)$ and $
20^2  (\bigtriangledown)$.    The error bars are smaller than the sizes of the symbols and the lines are guides to the eye.
\label{fig:6} }
\end{figure}

The effect of the use of  linearly Boolean separable input-output mappings on the measured performance of ACH can be appreciated in  Figure \ref{fig:6}  where we  show
 a comparison between the performance of that heuristic for the random and the linearly Boolean separable mapping. As mentioned before, in the case of the random mapping the minimum cost is not necessarily zero and the global minimum is obtained through an exhaustive search in the configuration space (hence the restriction to $F \leq 25$). Although the random mapping seems to be a harder problem to the  ACH, there is no qualitative difference between the dependence of our performance measure $P_m$  on the parameters $N$ and $F$ for the two mappings, and so our scaling results are likely to remain true  for the random mapping as well.

%----------------------------------------------------------------------------------------------------
%----------------------------------------------------------------------------------------------------

\begin{figure}
\includegraphics[width=0.52\textwidth]{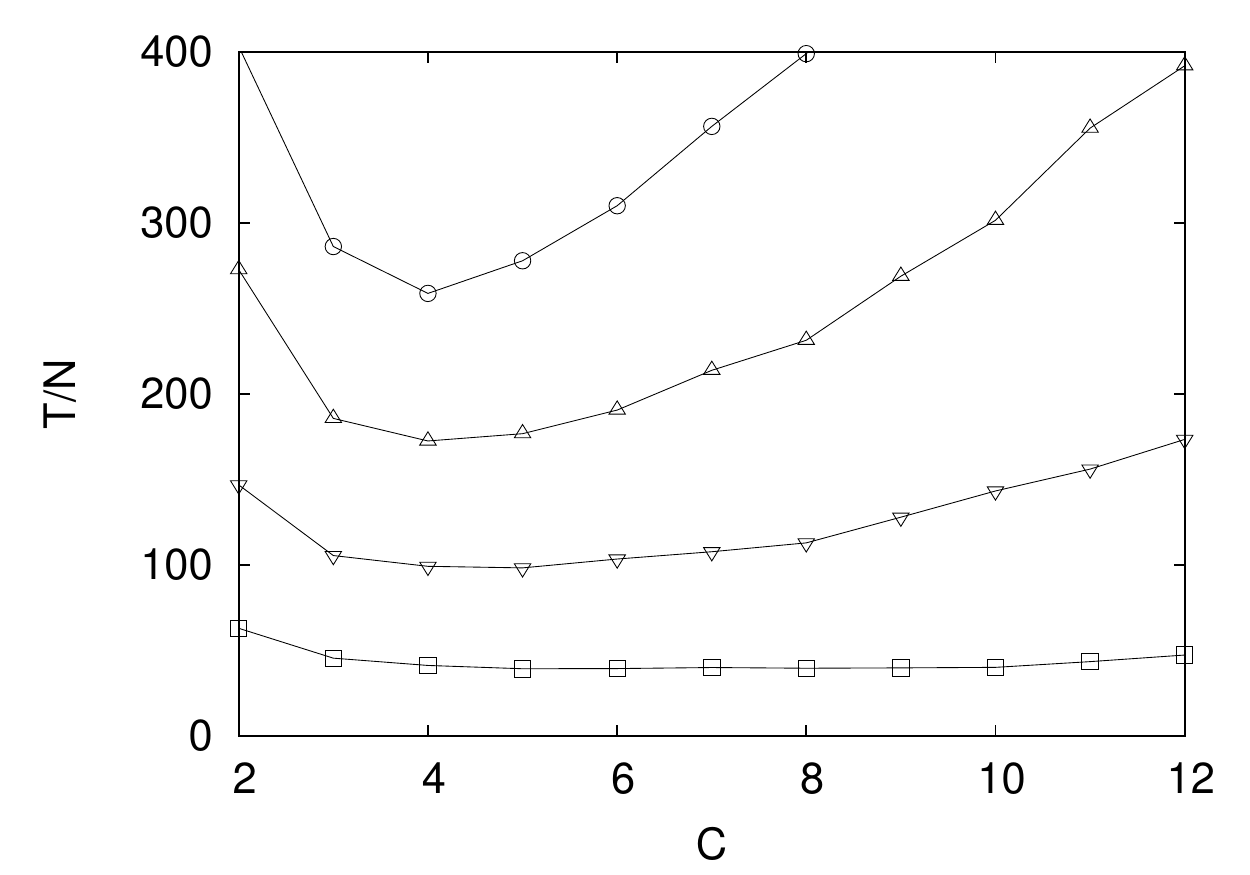}
%\centerline{\epsfig{width=0.52\textwidth,file=fig7.eps}}
\par  
\caption{Scaled relaxation time of the ACH as function of the connectivity $C$ of  random symmetric networks of $N=10^2$ agents for
$F= 11 (\square),  21 (\bigtriangledown), 31  (\triangle) $ and $ 41 (\bigcirc) $.  Each symbol represents the average over $10^3$ distinct random 
symmetric networks of fixed connectivity. The error bars are smaller than the sizes of the symbols and the lines are guides
to the eye. 
\label{fig:7} }
\end{figure}

To conclude our analysis, a word is in order about the impact of the connectivity between the agents on the performance of the ACH.  
It is well-known that the expansion of the influence range  of the agents,  modeled by 
increasing the connectivity of the lattice \cite{Greig_02,Klemm_03b} or by  placing the agents in more complex networks \cite{Klemm_03a}
(e.g., small-world and scale-free networks), results in the cultural homogenization of the population in Axelrod's model. Hence, it is not  unreasonable to
expect that by increasing the connectivity of the lattice (or network)  the relaxation time would decrease and so  the  computational cost of the heuristic 
would be reduced.  Alas, that is not so. In fact, the results of Fig.\ \ref{fig:7}, which  shows the scaled relaxation time $T/N$ as function of the 
connectivity $C$  of a random symmetric network composed of  $N=10^2$  agents,    indicate that $T/N$  reaches a minimum around $C=4$.
As expected,  we find that the probability $P_m$ of reaching the optimal solution is not affected by the  choice of the
connectivity  $C$,  and so the connectivity $C=4$ yields  the best performance, in the sense of the least computational cost,
of the ACH for not too small $F$. In addition,
the finding that the results of the random symmetric network with $C=4$ are indistinguishable from the results  obtained for
the regular square lattice (data not shown) suggests that the topology of the network does not influence the performance of the ACH.

%----------------------------------------------------------------------------------------------------
%----------------------------------------------------------------------------------------------------

\section{Conclusion}\label{sec:conc}

Understanding and quantifying how cooperation can improve the performance of groups of individuals to solve problems is an issue of great  interest to many areas - ranging from computer science to business administration    \cite{Clearwater_91}.
Our findings about the performance of the  Adaptive Culture Heuristic (ACH)  indicate that the number of agents participating of the collective solution of an optimization problem may influence the outcome of the process in a highly non-linear way (see, e.g., Fig.\ \ref{fig:1}). 

Our results were derived for a particular NP-Complete optimization problem, namely, the  classification of linearly Boolean separable input patterns by a  Boolean Binary  Perceptron,   whose optimal (zero-cost) solution is known by construction and which involves the manipulation of binary variables only. These two features allowed the 
study of the performance of the ACH for very large input sizes $F$ -- which essentially measures the `size' of the optimization problem -- and for a large number $N$ of agents  involved in the 
collective problem solving task. 

We focused on a single performance measure $P_m$,  which yields the probability that a run of the ACH  finds an optimal solution, and  found that it is a function of the reduced variable $u = F/N^ {1/4}$ for $N \geq 30$ (see Figs.\ \ref{fig:2} and \ref{fig:3}). This is a most remarkable and useful result which informs  how the number of  agents must scale with the problem size for a given fixed performance of the ACH, namely, $N \propto F^ 4$.  Recalling that the  scaled relaxation time $T/N$ scales with  $F^ 2$ (see Fig.\ \ref{fig:5}) we find that the overall computational cost
to find an optimal  solution with a fixed probability  scales with $F^ 6$.  As mentioned in Sect.\ \ref{sec:Intro}, this finding has no bearing on the $NP \neq P$ conjecture of computer science.
In addition, a surprising result, which is summarized in Fig.\ \ref{fig:7},  indicates that the implementation of the ACH  on a square lattice or on a random symmetric network of connectivity
$C=4$, yields the best performance when compared with the implementation on a random network of different connectivity.   

It  would be most interesting to find out whether the $F^ 6$ scaling law derived for the problem of learning linearly separable patterns by a Boolean Binary Perceptron holds for other optimization problems 
as well. In that case, one would have revealed  a genuine  property  of the ACH which, given the minimal nature of the underlying social interaction mechanism, might serve as a bound to
the performance of heuristics based on collective computation.

\begin{acknowledgments}
This research   was  supported by  The Southern Office of Aerospace Research and Development (SOARD), grant FA9550-10-1-0006,  
and Conselho Nacional de Desenvolvimento Cient\'{\i}fico e Tecnol\'ogico (CNPq).
\end{acknowledgments}

% References

\end{document}